\documentclass[aps,preprint,superscriptaddress]{revtex4-1}
\usepackage{graphicx}
\usepackage{amsmath}
\usepackage{mathtools}
\usepackage{amssymb}
\usepackage[dvipdfm]{hyperref}

\begin{document}

\title{Asymmetric magnetic bubble expansion under in-plane field in Pt/Co/Pt: effect of interface engineering}

\author{R.~Lavrijsen}
\email[]{r.lavrijsen@tue.nl}
\affiliation{Department of Applied Physics, Center for NanoMaterials and COBRA Research Institute, Eindhoven University of Technology, P.O.~Box~513, 5600 MB Eindhoven, The Netherlands}
\author{D.M.F. Hartmann}
\affiliation{Department of Applied Physics, Center for NanoMaterials and COBRA Research Institute, Eindhoven University of Technology, P.O.~Box~513, 5600 MB Eindhoven, The Netherlands}
\affiliation{Institute for Theoretical Physics, Utrecht University, Leuvenlaan 4, NL-3584 CE Utrecht, The Netherlands}
\author{A. van den Brink}
\author{Y. Yin}
\author{B. Barcones}
\affiliation{Department of Applied Physics, Center for NanoMaterials and COBRA Research Institute, Eindhoven University of Technology, P.O.~Box~513, 5600 MB Eindhoven, The Netherlands}
\author{R.A. Duine}
\affiliation{Institute for Theoretical Physics, Utrecht University, Leuvenlaan 4, NL-3584 CE Utrecht, The Netherlands}
\author{M.A. Verheijen}
\affiliation{Department of Applied Physics, Center for NanoMaterials and COBRA Research Institute, Eindhoven University of Technology, P.O.~Box~513, 5600 MB Eindhoven, The Netherlands}
\affiliation{MiPlaza Technology Laboratories, Philips Research Europe, High Tech Campus 29, 5656 AE Eindhoven, The Netherlands}
\author{H.J.M. Swagten}
\author{B. Koopmans}
\affiliation{Department of Applied Physics, Center for NanoMaterials and COBRA Research Institute, Eindhoven University of Technology, P.O.~Box~513, 5600 MB Eindhoven, The Netherlands}

\date{\today}

\begin{abstract}
We analyse the impact of growth conditions on asymmetric magnetic bubble expansion under in-plane field in ultrathin Pt / Co / Pt films. Specifically, using sputter deposition we vary the Ar pressure during the growth of the top Pt layer. This induces a large change in the interfacial structure as evidenced by a factor three change in the effective perpendicular magnetic anisotropy. Strikingly, a discrepancy between the current theory for domain-wall propagation based on a simple domain-wall energy density and our experimental results is found. This calls for further theoretical development of domain-wall creep under in-plane fields and varying structural asymmetry.
\end{abstract}

\maketitle

\subsection{Introduction}\label{SEC:INTRO}
The recent manifestation of interfacial Dzyaloshinskii-Moriya interaction (DMI) \cite{DZYALOSHINSKY1958,MORIYA1960,FERT1980} in nominally symmetric ultrathin Pt / Co / Pt films and Pt / [Co / Ni]$_N$ multilayers \cite{JE2013,HRABEC2014,FRANKEN2014,EMORI2013,RYU2013} has raised questions concerning the origin of this interaction. In these multilayers DMI is usually explained in terms of structural inversion asymmetry, that arises as a result of the asymetric stacking of materials, hence, in symmetric Pt / Co / Pt the structural asymmetry should have a different origin. Most likely the effective DMI that is found in these structures arises from an asymmetry in the interfaces at either side of the ultrathin ferromagnetic layer \cite{BODE2007,HRABEC2014,CHEN2013,HEIDE2008,CHEN2013a}. For Pt / Co / Pt samples even opposite signs of the DMI induced fields have been reported, where Je \textit{et al.} \cite{JE2013} found an DMI induced field of $\mu_0H_{DMI} \approx$ +26 mT, Hrabec \textit{et al.} \cite{HRABEC2014} found $\mu_0H_{DMI} \approx$ -100 mT. Even more striking is that when replacing the top Pt layer by an Ir layer a reversal in $H_{DMI}$ is observed. This is counterintuitive as Pt / Co and Ir / Co interfaces are expected to have opposite DMI \cite{CHEN2013,RYU2013}, which should effectively increase the net $H_{DMI}$ in a Pt / Co / Ir multilayer. The origin of these contradicting reports might lie in interfacial quality-defining properties such as roughness, degree of intermixing, etc. Characterizing these structural properties quantitatively however, remains an outstanding challenge due to the ultrathin ferromagnetic layers used (typically $<$ 1 nm) and poly-crystalline nature of the, typically sputter-deposited, films. Furthermore, in the area of magnetic field and current-induced domain-wall (DW) motion there have been widely differing reports on the strength of the DMI \cite{FRANKEN2014,EMORI2013,RYU2013,THIAVILLE2012}. This might be due to intrinsic differences between the deposited films grown in different laboratories. Moreover, any comparison is hampered since parameters such as growth rate, power, gas pressure and sample-substrate details are missing and, if given, are difficult to compare as the used specific sputter apparatus and preparation method are also defining. Tackling these issues are key to understanding, comparing and interpreting the reported DMI results. The urgency of understanding the DMI in ultrathin films arises from the predicted huge impact on technological relevant devices where a chiral magnetization texture is preferred such as racetrack memories and logic devices based on DWs and skyrmions \cite{PARKIN2008,EMORI2013,BOULLE2013,SAMPAIO2013,DUPE2014,STAMPS2014}.

Here, we investigate the effect of growth conditions on the DMI by means of magnetic bubble expansion under in-plane fields in the archetype films of nominally symmetric Pt(4 nm) / Co(0.6 nm) / Pt(4 nm) films as was successfully applied before \cite{JE2013,HRABEC2014}. Specifically, using DC magnetron sputter deposition, we vary the Ar gas pressure during the growth of the top Pt layer $p_{top}$. Thereby, we change the growth kinetics of the top Pt layer \cite{BERTERO1994,BERTERO1995}. This leads to a different interfacial quality and/or degree of intermixing, and hence a variation in the degree of structural inversion asymmetry, giving rise to an effective interfacial DMI interaction \cite{BODE2007}, and magnetic properties. We find a large dependence of the perpendicular magnetic anisotropy (PMA) on $p_{top}$, varying up to a factor three between the lowest and highest $p_{top}$. Asymmetric bubble expansion under applied in-plane fields is observed indicating a finite DMI \cite{JE2013,HRABEC2014}. We are, however, unable to describe our experimental data with the simple theory used before \cite{JE2013,HRABEC2014}. This indicates a more complicated physical picture and the currently used models need to be expanded to incorporate the complex behavior observed. Our results shed light on the origin of the interfacial DMI in sputter deposited Pt / Co / Pt layers and provide a simple way to investigate the effect of changing interfacial quality.

This letter is structured in the following order; we will start with introducing the used experimental methods in section \ref{SEC:METHODS}. In section \ref{SEC:PMA} we will present the basic magnetic properties of our samples. Here the large effect of the Ar pressure during the top Pt layer growth on the PMA is shown. In section \ref{SEC:CREEP} the results of expanding bubbles are presented, furthermore, we will confirm that the DWs in all our samples follow the creep law. Furthermore, we will extract the creep law parameters and correlate this with the behavior of PMA as a function of $p_{top}$. In section \ref{SEC:INPLANE} we will concentrate on the results of bubble expansion under in-plane fields. We will elaborate on the differences and correspondence of the data with the current understanding. Finally, we will discuss the results and conclude in section \ref{SEC:CONCL}.

\subsection{Methods}\label{SEC:METHODS}
The samples are grown using parallel face-to-face target and substrate Ar DC magnetron sputter deposition from 2" targets in a system with a base pressure of $\sim3\times10^{-8}$ mbar. All samples are prepared on precut Si substrates (0.5x0.5 cm$^2$) with a native oxide layer. The substrates were cleaned by acetone and isopropanol in an ultrasound bath followed by an in-situ 5W O$_2$ plasma exposure for 10 minutes prior to the deposition. The studied samples consist of SiO2//Ta(4)/Pt$_{\textrm{L}}$(4)/Co(0.6)/Pt$_{\textrm{U}}$(4) (thickness between parenthesis is given in nm) where we have labeled the lower and upper Pt layers as Pt$_{\textrm{L}}$ and Pt$_{\textrm{U}}$. For Ta, Co and Pt DC magnetron powers of 20, 20, 60 W were used, respectively. The target-sample distance during sputtering for Ta, Co and Pt was kept constant at 95, 95 and 195 mm, respectively. The Ar pressure during Ta, Co and Pt$_{\textrm{L}}$ layer deposition was kept constant at 1.4, 1.0 and 0.29 Pa, respectively. The pressure $p_{top}$ during the Pt$_{\textrm{u}}$ layers was varied between 0.29 and 2.8 Pa. The Pt growth rate was calibrated for every pressure to keep the top Pt layer thickness constant at 4 nm. The saturation magnetization was measured using a SQUID-VSM at room temperature. The perpendicular magnetic anisotropy was determined using angle dependent anomalous Hall effect measurements at room temperature. A standard polar MOKE setup was used to obtain the hysteresis loops at a constant field sweep rate of 10 mT/s. The bubble expansion measurements were performed using wide field Kerr microscopy in polar mode with a custom perpendicular pulse coil with a calibrated rise time of 70 $\mu$s to provide the perpendicular to the surface z-fields and a standard in-plane magnet for the in-plane x-field.

\subsection{Basic magnetic properties as a function of $p_{top}$}\label{SEC:PMA}
\begin{figure}[p]
     \begin{center}
      \includegraphics[width=1.0\textwidth]{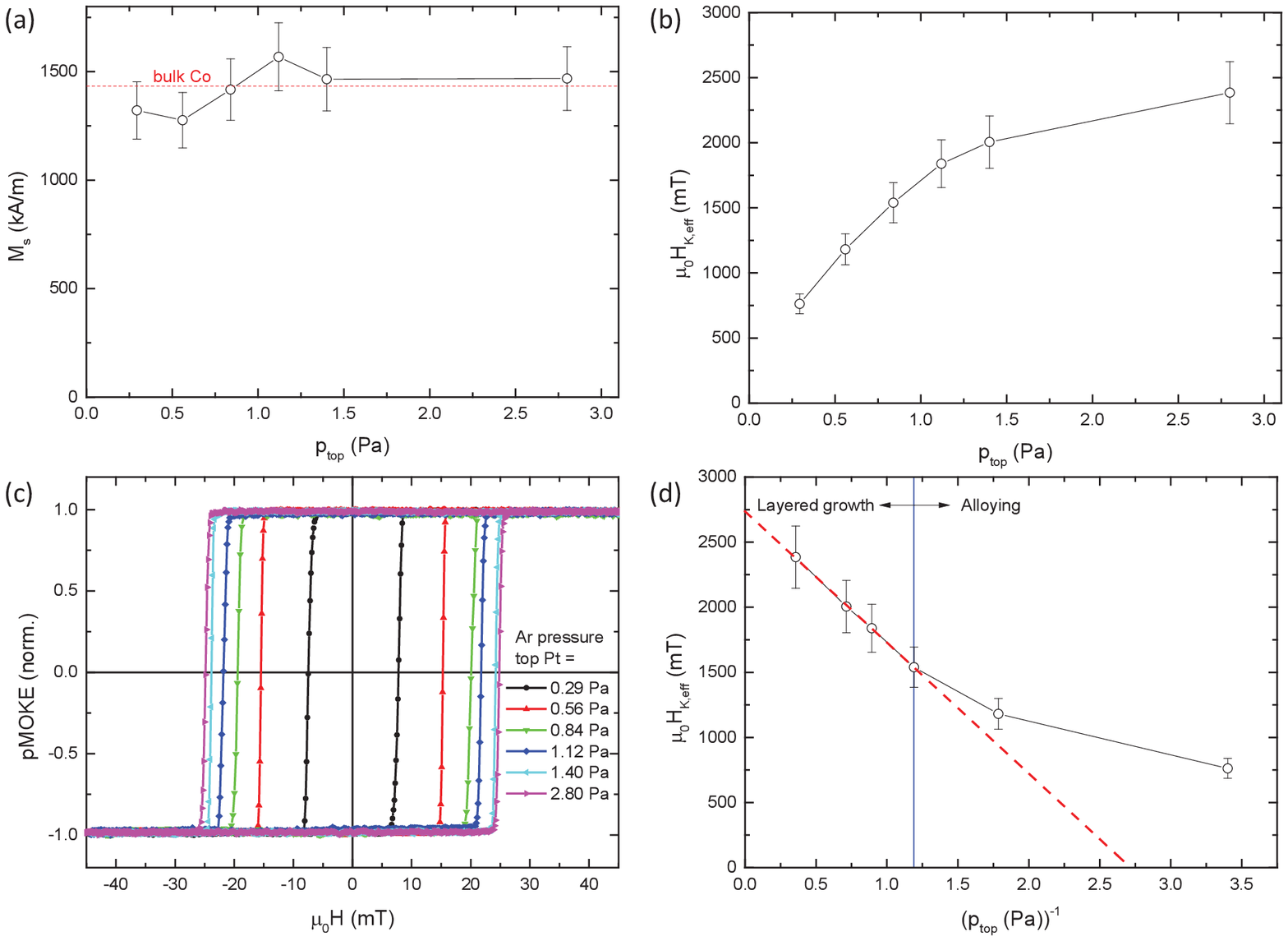}
      \caption{(a) Saturation magnetization $M_S$ as a function of $p_{top}$. (b) Effective anisotropy field $\mu_0H_{K,eff}$ as function of $p_{top}$. (c) Polar MOKE loops for different $p_{top}$. (d) Effective anisotropy field $\mu_0H_{K,eff}$ as a function of $p_{top}^{-1}$, the red dash line is a linear fit to the data for $p_{top}^{-1}<1.19$ Pa$^{-1}$. The blue line indicates a transitions between the `layered growth' and `alloying' regime.}
      \label{PMA}
    \end{center}
    \vspace{-3mm}
\end{figure}

In Fig. \ref{PMA}(a) the measured saturation magnetization $M_S$ as a function $p_{top}$ is plotted for SiO$_2$ // Ta(4 nm) / Pt(4 nm) / Co(0.6 nm) / Pt(4 nm). The $M_S$ is found to be close to the bulk value of Co (1440 kA/m) for all used $p_{top}$. In Fig. \ref{PMA}(b) we plot the effective anisotropy field $H_{K,eff}$ as a function of $p_{top}$ which shows a clear increase with $p_{top}$ up to a factor 3.1 between $p_{top}$ = 0.29 Pa (760$\pm$80 mT) and $p_{top}$ = 2.80 Pa (2380$\pm$240 mT).

In Fig. \ref{PMA}(c) the hysteresis loops of the samples are presented showing square loops and increasing coercivity $H_c$ for higher $p_{top}$, in agreement with observed behavior of $H_{K,eff}$. These results show a clear trend of the perpendicular magnetic anisotropy (PMA) as captured in $H_{K,eff}$ with increasing $p_{top}$ illustrating the sensitive nature of the PMA to the growth conditions. Hence, the structural quality of the Co layer and interfaces with the Pt are paramount and are determined by the kinetics during the growth of the top Pt layer \cite{BERTERO1994,BERTERO1995}. Similar behavior has been seen before where the PMA was fully dominated by the bottom Pt / Co interface \cite{HRABEC2014}. In another report the lack of PMA from the top Co / Pt was attributed to a strongly intermixed top Co / Pt interface \cite{BANDIERA2011}. Hence, extending on this scenario a basic understanding can be found by examining the incoming kinetic energy of the Pt atoms during growth, which roughly scales as the inverse of the pressure \cite{BERTERO1994,BERTERO1995}. To test this we plot in Fig. \ref{PMA}(d) $H_{K,eff}$ as a function of $p_{top}^{-1}$. Indeed for high pressure ($>$ 0.8 Pa, $<$ 1.19 Pa$^{-1}$) a linear scaling with $p_{top}^{-1}$ is found as indicated by the linear fit (red dashed line). For lower $p_{top}$ this trend breaks down. We attribute this break down as the region where there is a strong intermixing between the Pt and Co atoms possibly creating an alloyed PtCo layer, which are known to show high PMA. In this case no well-defined Pt / Co and Co / Pt interfaces are formed, i.e. an interfacial structural inversion asymmetry would be poorly defined.

For higher pressures there is less intermixing and the interfaces get increasingly better defined. In this regime an effective DMI due to interfacial structural inversion asymmetry could be expected as long as the bottom and top interface are sufficiently different. We have labeled these regions the `alloying' and `layered growth' regime, respectively, in Fig. \ref{PMA}(d). Other origins for the increasing PMA with $p_{top}$ could be a changing density of the top Pt film with $p_{top}$, different crystal orientations, growth modes and/or grain sizes.

We have performed cross-sectional and plain view high resolution transmission electron microscopy (HRTEM) studies on samples prepared on SiN$_x$ membranes and 50 nm thick lamella prepared using a focused ion beam but were unable to discern differences between the samples grown with low and high $p_{top}$, both in crystal size as in orientation, viz. a strong $<$111$>$ texture for all samples was observed from electron diffraction patterns. Moreover, using scanning TEM - EDS (energy dispersive X-ray Spectroscopy) mapping we were unable to discern any differences in the intermixing between Pt and Co. This is due to the limit in resolution in the element maps due to the combination of the limited thickness of the ultrathin Co film (0.6 nm) and the surface roughness of the underlying Pt layer (due to its poly-crystalline nature, yielding a projected layer thickness of the Co layer of $\sim$1.5 nm. Future atom-probe tomography measurements might shed light on the structural properties and provide further proof for the given hypothesis. We conclude that increasing the Ar pressure during the growth of the top Pt layer in Pt / Co / Pt layers increases the effective PMA by a factor 3.1. We attribute this change to an increasingly better defined layered growth of the Pt / Co / Pt stack.

\subsection{DW creep - bubble expansion}\label{SEC:CREEP}

\begin{figure}[p]
     \begin{center}
      \includegraphics[width=0.7\textwidth]{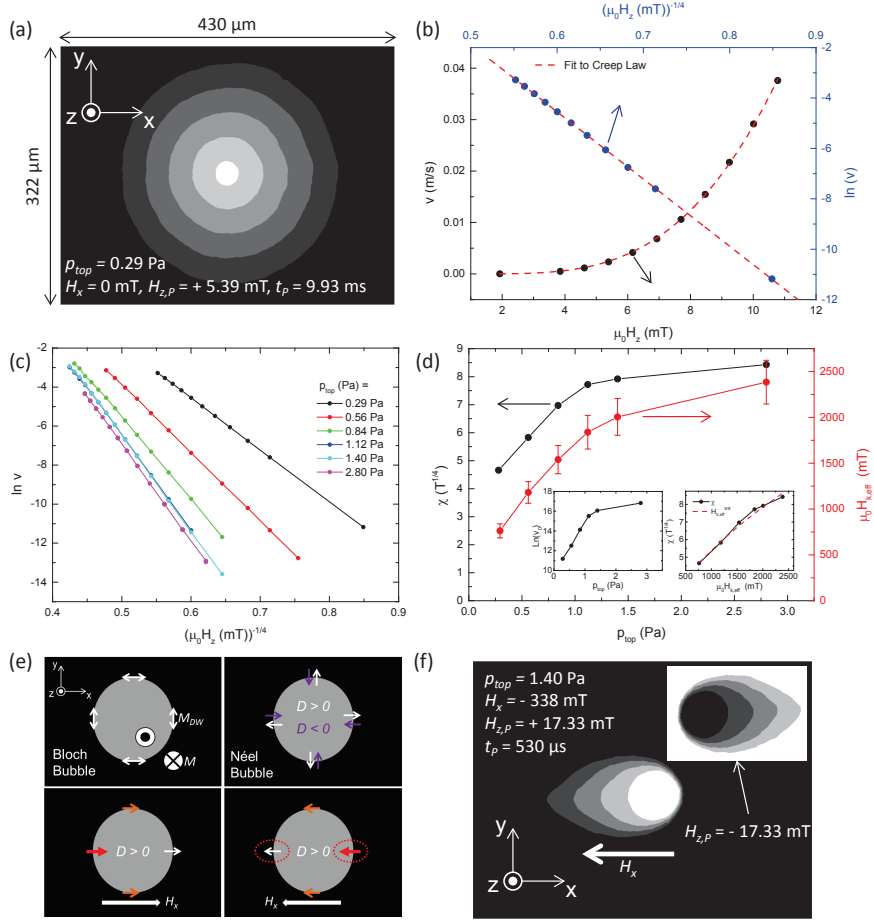}
      \caption{(a) Typical example of a symmetrically expanding bubble with $H_x$ = 0 mT driven by an out-of-plane magnetic field $H_{z,P}$ of 5.39 mT with a duration of $t_{P}$ = 9.93 ms for the $p_{top}$ = 0.29 Pa sample. (b) Typical velocity $v$ versus $\mu_0H_z$, black symbols and left-bottom axis and $\ln(v)$ versus $(\mu_0H_z)^{-1/4}$, blue symbols and right-top axis for $p_{top}$ = 0.29 Pa. The red dashed lines are a fit to the creep law. (c) $\ln(v)$ versus $(\mu_0H_z)^{-1/4}$ for all samples. (d) Creep law scaling constant $\chi$ versus $p_{top}$ (black symbols and right/bottom axis). Effective anisotropy field as function of $p_{top}$ (red symbols and right/bottom axis). The left inset shows the creep law prefactor $v_0$ as a function of $p_{top}$. The right inset plot $\chi$ versus $K_{eff}$ indicating the $\chi \sim K_{eff}^{5/8}$ proportionality. (e) Top left panel shows a Bloch bubble where we have indicated the magnetization angle inside the DW defining the bubble. The top right panel shows a N\'{e}el type bubble where the chirality of the DW depends on the sign of the DMI interaction. Bottom two panels show the orientation when the DWs magnetization is saturated along the applied in-plane field direction and the thickness of the arrows indicate the increase in energy of the DWs for different $D$, the white dotted circles in the bottom right panel accentuate the change when the in-plane field is reversed. (f) Bubble expansion (same scaling as in (a)) in the $p_{top}$ = 1.40 Pa grown sample while applying an in-plane field, the inset shows the expansion with inverted $H_{z}$ showing identical, but mirrored in the y-axis, expansion.}
      \label{BUB}
    \end{center}
    \vspace{-3mm}
\end{figure}

In Fig. \ref{BUB}(a) a typical example of a symmetrically expanding bubble is shown under application of $H_z$ field pulses using the following measurement sequence (identical expansion is observed for all samples). (1) The sample was saturated with a z-field $H_{z,S}$ pulse with a length of $t_{S}$. (2) A background image is taken and subtracted. (3) A circular bubble is then nucleated using a nucleation pulse ($H_{z,N}$,$t_{N}$). (4) The in-plane $H_x$ field was set. (5) An initial reference image is taken. (6a) A propagating pulse ($H_{z,P}$,$t_{P}$) is applied. (6b) An image is recorded. Step 6 is then repeated $X$ times. (7) $H_x$ is switched off. The obtained images are then processed to extract the bubble nucleation origin and DW velocity along the bubble circumference. Special care was taken to find a nucleation site where a consistent and circular bubble was nucleated with both polarities during the nucleation pulse. Furthermore, the area around the nucleation site, i.e. the area where the expanding bubble travels through, was checked for strong pinning sites which could possibly lead to non-continuous extracted DW velocities. After mounting every sample the in-plane field coil angle relative to the sample plane was carefully tuned to minimize the leak into the perpendicular field direction.

By fitting and averaging the expansion of the bubble versus the total applied $H_z$ pulse time the DW velocity $v$ versus $\mu_0H_z$ is extracted, as shown in Fig. \ref{BUB}(b) (black symbols). The data can be fitted to the creep law \cite{LEMERLE1998}:
 \begin{equation}\label{CREEP}
    v = v_0\exp(-\chi(\mu_0H_z)^{-1/4}).
\end{equation}
The red dashed lines in Fig. \ref{BUB}(b) are fits to the creep law and an excellent agreement is found. The blue symbols correspond to the top and right axis where we have plotted $\ln(v)$ versus ($\mu_0H_z)^{-1/4}$ showing a linear behavior, as expected. Identical behavior is observed in all samples as is shown in Fig. \ref{BUB}(c). In Fig. \ref{BUB}(d) we plot the characteristic fit parameters $v_0$ (left inset) and $\chi$ as a function of $p_{top}$ obtained from fits to the data shown in Fig. \ref{BUB}(b). Furthermore, $v_0$ is the characteristic speed parameter and scales with $H_{K,eff}$. In the creep description $v_0$ is defined as the product of the correlation length $\xi$ of the disorder potential (average distance between pinning sites) and the depinning frequency $f_0$ \cite{CHAUVE00}. From this data, however, we cannot distinguish which of these two parameters changes as a function of $p_{top}$. Speculative, we expect the change in $\xi$ with $p_{top}$ to be small as we observe identical structural properties for all $p_{top}$ from the HRTEM study and the strong increase in $v_0$ could be dominated by $f_0$, this however, remains an open issue. To fully understand this relation further theoretical investigation is needed.

The scaling constant $\chi$, scales with $H_{K,eff}$ which is plotted in red in Fig. \ref{BUB}(d) (right axis). $\chi$ is originally defined as $U_cH_{crit}^{\mu}/(k_BT)$ in the DW creep theory with $\mu = 1/4$ \cite{LEMERLE1998,KIM2009}, where $U_c$ is an energy scaling constant, $H_{crit}^{\mu}$ is the critical magnetic depinning field at zero temperature, and $k_BT$ denotes the thermal energy. Using the description of the creep law $\chi \sim H_{K,eff}^{5/8}$ can be obtained \cite{KIM2014}. In the right inset of Fig. \ref{BUB}(d) we have plotted $\chi$ versus $H_{K,eff}$ and the dashed red line shows $\chi \sim H_{K,eff}^{5/8}$. The good agreement leads us to conclude that no large changes in the shape of the DW pinning landscape is expected, as it only scales with $H_{K,eff}$. Summarizing, we conclude that the DW motion is well characterized by the DW creep law for all $p_{top}$ and $H_z$, i.e. velocities, studied.

\subsection{Bubble expansion under in-plane field}\label{SEC:INPLANE}

In Pt / Co / Pt thin films with PMA, Bloch type DWs are preferred due to the magneto-static DW anisotropy field $H_{K,dw} = 4K_{dw}/(\pi\mu_0M_S)$ with $K_{dw} = N_x\mu_0M_S^2/2$ the magneto-static DW anisotropy, where $N_x = t_{Co}\log(2)/(\pi\Delta)$ is the demagnetization coefficient of the DW \cite{TARASENKO1998} with $t_{Co}$ the Co layer thickness and $M_S$ the saturation magnetization. For magnetic bubbles this leads to the situation as shown in the top left panel of Fig. \ref{BUB}(e) where a magnetic bubble with its magnetization along $+z$ (grey area) is shown. Since there is no preferred chirality for Bloch DWs the DW magnetization can rotate either clock or anti-clockwise going through the DW from the inside to the outside of the bubble as indicated by the double arrows. In a simple description of DMI at DWs \cite{CHEN2013,THIAVILLE2012,JE2013,HRABEC2014}, the DMI manifests itself as a built-in magnetic field $H_{DMI} = D/(\mu_0M_S\Delta)$ pointing in-plane perpendicular to the DW, where $D$ is the strength of the DMI interaction, and $\Delta$ the DW width. Hence, for $H_{DMI} > H_{K,dw}$ N\'{e}el type DWs are preferred as shown in the top right panel of Fig. \ref{BUB}(e) and depending on the sign of $D$ a clock ($D > 0$), or anti-clockwise ($D < 0$) DW chirality is introduced as indicated by the white and purple arrows, respectively. For $H_{DMI} < H_{K,dw}$ the DW assumes a mixed Bloch-N\'{e}el character. By now applying a strong in-plane field along the x-axis ($H_x$) the DWs magnetization reorients itself along $H_x$ due to the Zeeman energy. This is shown for a bubble with $D > 0$ in the bottom left panel of Fig. \ref{BUB}(e), where we assume $H_{DMI} > H_{K,dw}$. The DW segments that are parallel to $H_x$ (top and bottom of bubble, orange arrows) become Bloch type DWs with an increase in DW energy density due the Zeeman energy. The DW segments that are perpendicular (left and right of bubble) remain of the N\'{e}el type. The N\'{e}el DW which has its direction reversed (and thus its chirality) however, undergoes an increase in DW energy density. The same happens when $D > 0$ (bottom right panel of Fig. \ref{BUB}(e)), albeit here the left and right DW behavior are reversed. Hence, an in-plane magnetic field breaks the symmetry of the DW energy profile along the in-plane field direction.

This is experimentally shown in Fig. \ref{BUB}(f) where the bubble expansion for the $p_{top}$ = 1.40 Pa sample is shown with a $H_x$ field applied. A strongly asymmetric expansion is observed where the DW moving in the direction of the applied $H_x$ field moves much faster as the DW moving against the in-plane field. The inset shows the expansion in the same sample but with inverted $H_z$ showing identical asymmetric expansion albeit mirrored in the y-axis. This observation was attributed to a built-in DMI field manifesting itself at the DWs as explained before \cite{JE2013,HRABEC2014}. By applying an in-plane field during the expansion the radial symmetry is broken as the DMI field would prefer N\'{e}el type DWs with a certain chirality which is broken by the in-plane field.

\begin{figure}[p]
     \begin{center}
      \includegraphics[width=0.9\textwidth]{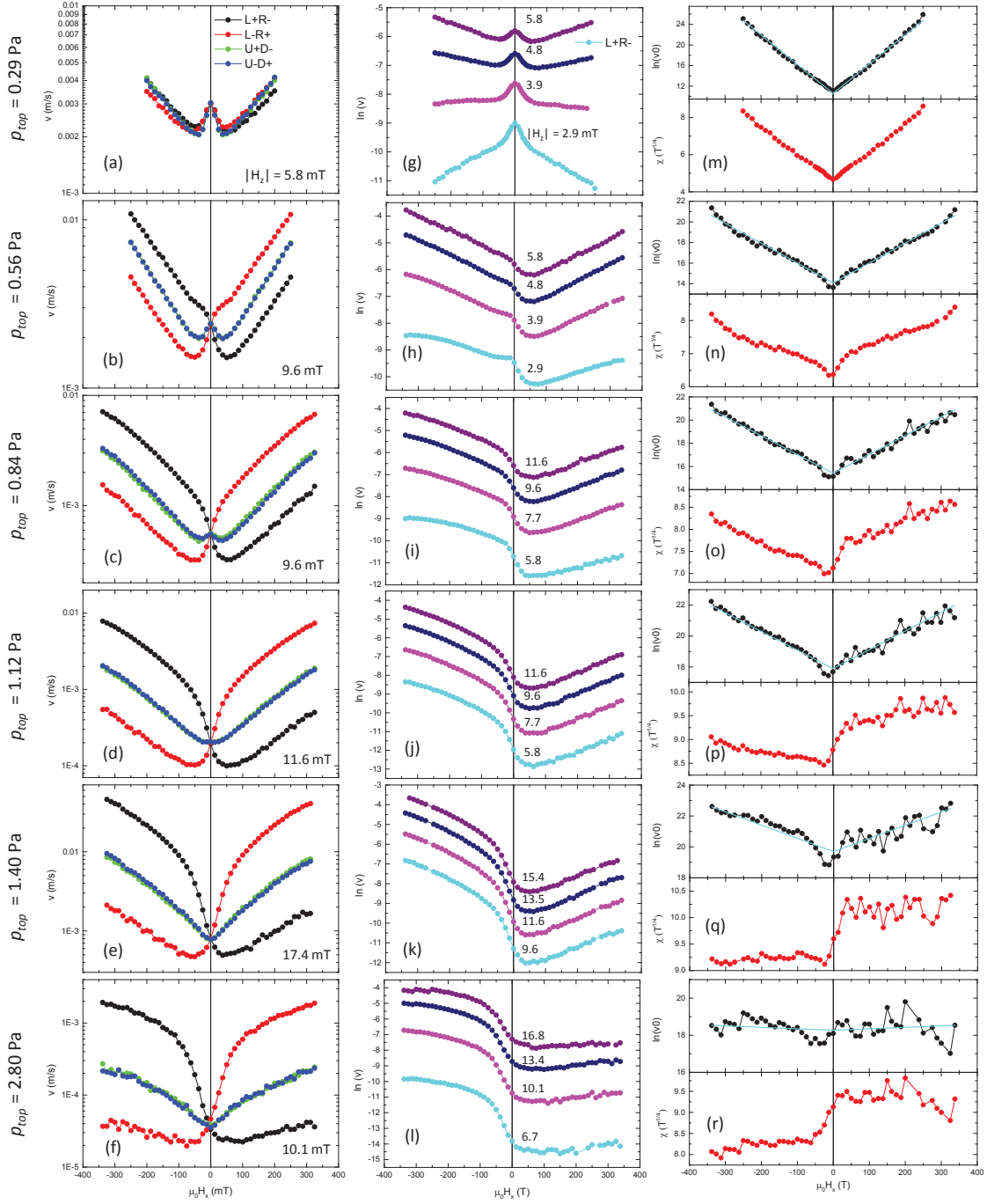}
      \caption{(a - f) DW velocity profiles for L+R-, L-R+, U+D- and U-D+ on a log scale as a function of $H_x$ for all $p_{top}$. Note the different ranges of the $v$ scale used, i.e. the velocity variation increases with increasing pressure $p_{top}$. (g)-(l) L+R- velocity profiles as a function of $H_x$ for all $p_{top}$ and different $|H_z|$ as indicated in the panels. (m)-(r) $\ln(v_0)$ and $\chi$ as a function of $H_x$ for all $p_{top}$ as extracted from the data shown in (g)-(l). The cyan line in the $\ln{v_0}$ graph indicates the symmetric behavior relative to $H_x$ = 0.  }
      \label{OVERVIEW}
    \end{center}
    \vspace{-3mm}
\end{figure}

In Fig. \ref{OVERVIEW}(a)-(f) we plot the obtained DW velocities $v$ as a function of $H_x$ along the extremes of the bubble along the x and y-axis, i.e. parallel and perpendicular to the applied $|H_x|$ field for all $p_{top}$. Note the different $v$ scales used; the variation in $v$ increases with higher $p_{top}$. We have averaged the left moving DW segment velocity with positive $H_{z} > 0$ (i.e. up-down DWs when moving from inside the bubble to outside), which we label L+, with the right moving DW velocity for $H_{z} < 0$ (R-) (i.e. down-up DWs when moving from inside the bubble to outside), as these have to be identical by symmetry (see Fig.\ref{BUB}(f)).

The same has been done for the up (U) and down (D) moving DW segments. This allows us to compensate for small misalignments of the, relative to $H_z$, large $H_x$ field. Despite this compensation small residual asymmetries remain which we consider negligible (compare the U+D- (green) and U-D+ (blue) profiles which need to be similar by symmetry). Extracting the DW velocity profiles in this way allows us to disentangle the in-plane field effect on the DW velocity parallel (L+R- and L-R+) and perpendicular (U+D- and U-D+) to the applied $H_x$ field. The asymmetric bubble expansion is reflected in the difference between the L+R- and L-R+ velocity profiles at a certain $H_x$. The U+D- and U-D+ profiles show a symmetric profile around $H_x$ = 0 mT. This observation corresponds with the aforementioned picture of a DMI field breaking the in-plane symmetry of the bubble expansion.

For $p_{top}$ = 0.29 Pa an increasingly ill-defined bubble expansion and extra domain nucleation is observed for high $|H_x|$ (not shown). This makes it impossible to extract well defined DW velocities for $|H_x| >$ 200 mT as there is no bubble expansion but a dendrite like magnetization reversal. On close inspection of all the profiles three distinct DW velocity profile characteristics can be identified: (1) symmetric bubble expansion perpendicular to the in-plane field direction (U+D- and U-D+). (2) Asymmetric bubble expansion parallel to the in-plane field direction (L+R- and L-R+). (3) For $p_{top}$ = 0.29 Pa the DW velocity shows a velocity drop symmetrically around $H_x$ = 0 (most clearly seen for the U+D- and U-D+ profiles) which decreases for the samples grown with higher $p_{top}$.

All the experimental L+R- (black symbols) profiles show a minimum at $H_x \approx 60$ mT. In the interpretation of an effective DMI field this would indicate a constant positive built-in DMI field $H_{DMI} > 0$, i.e. a positive DMI energy $D > 0$ would be found for all samples preferring a right handed DW chirality. Furthermore, there would be no variation in the DMI as a function of $p_{top}$ which is puzzeling as the large variation in PMA indicates a strong change in the interfacial quality and speculative also on the structural inversion asymmetry. The found $D > 0$ is in-line with the report of Je \textit{et al.} but opposite to the report of Hrabec \textit{et al.} for Pt / Co / Pt. Moreover, the overall $v$ profile shape observed here is very different than described by the theory used in these reports. This leads us to conclude that the interpretation of the DMI leading to a simple in-plane field might be too simple for our samples.

The out-of-plane driving field might have an effect on the profile shapes as this determines the strength of the DW motion. To examine the effect of the out-of-plane driving field we have plotted the $\ln{v}$ L+R- profiles in Fig. \ref{OVERVIEW}(g)-(l) for different $|H_z|$ drive fields. For the lowest $p_{top}$ we observe an overall changing shape of the velocity profile with increasing $|H_z|$, for the higher $p_{top}$ the overall shape remains constant. This behavior further hints towards different mechanisms at work between the low and high $p_{top}$ samples which might be due to different DMI strengths. From this data we can extract the $v_0$ and $\chi$ parameters by fitting the velocity dependence on $|H_z|$ to the creep law (Eq. \ref{CREEP}) for every $H_x$. This is shown in Fig. \ref{OVERVIEW}(m)-(r) where we plot $\ln(v_0)$ and $\chi$ as a function of $H_x$. The cyan line in the $\ln(v_0)$ data is a fit to $\ln(v_0) = a + b|H_x|$, which shows that a symmetric behavior for $v_0$ is observed relative to $H_x$ = 0 for all samples. Furthermore, the variation in $\ln(v_0)$ decreases strongly with increasing $p_{top}$ and remains more or less constant for the highest $p_{top}$, which indicates a changing underlying dynamic response captured in $v_0$. The $\chi$ data, however, shows an increasing asymmetry relative to $H_x$ = 0 with higher $p_{top}$. The origin of this effect is not clear at the moment but the overal shape of the $\chi(H_x)$ dependence is very different as expected from the theory described by Je \textit{et al.}, where a simple inverted parabolic behavior is expected with a maximum in $\chi$ at $-H_{DMI}$.

\subsection{Discussion \& Conclusion}\label{SEC:CONCL}
Due to the discrepancy between the previously used theory and our experimental data we are unable to fully interpret the experimentally obtained velocity profiles. Naively interpreting the $H_x$ field at which a minimum in the DW velocity was observed, we find $\mu_0H_{DMI} \approx$  60$\pm$10 mT for all $p_{top}$. If we interpret the interfacial $H_{DMI}$ as an effective field, i.e. originating from the difference of $D$ between the bottom Pt / Co and top Co / Pt interfaces, a correlation between the PMA (through $p_{top}$) and $H_{DMI}$ should have been evident. This is motivated by the known dependence of the PMA on the interface quality. As this is not the case, i.e. a constant $H_{DMI}$ is found for all $p_{top}$ whilst the PMA varies over a factor three, indicates that this interpretation might be flawed. Hence, this calls for further theoretical development of the underlying mechanisms and the simple model introduced by Je \textit{et al.} should be used with great care. Specifically, we suggest that full micro-magnetics of the DWs profiles and the strong pinning of the DWs in the creep regime should be considered. The successful application of the theory in the reports by Je \textit{et al.}\cite{JE2013} and Hrabec \textit{et al.}\cite{HRABEC2014} might be due to different preparation procedures and (lower) $H_x$ field regimes studied. The Co layer in our samples are rather thick (0.6 nm) compared to the sample used by Je \textit{et al.}\cite{JE2013} (0.3 nm) which exhibit ultra low DW pinning, but are similar to the samples grown by Hrabec \textit{et al.}\cite{HRABEC2014}.

Very recently, asymmetric bubble expansion under in-plane fields was attributed to a completely different mechanism, viz. chiral-dependent damping \cite{MIHAI2014} again in similar systems of Pt / Co / Pt. This would modulate the attempt frequency $v_0$ in the creep law (Eq. \ref{CREEP}) as a function of in-plane field and was explained as a dissipative (field-like) spin-orbit torque on the DW dynamics. As we have seen from the data in Fig. \ref{OVERVIEW}(m)-(r) we observe a symmetric behavior of $v_0$ relative to $H_x$ = 0 and can exclude this in our samples. In fact, the experimental finding that $\chi$, and not $\ln v_0$, is asymmetric with respect to in-plane field suggests that the main asymmetry is in the DW energy rather than its dissipation. This conclusion holds provided the collective-pinning theory of the creep regime applies but is otherwise model-independent. One possibility for the influence of in-plane fields on the DW energy not considered before is the tilting of the domains by the in-plane field. This might be significant as the in-plane fields we apply are not necessarily small compared to the PMA. The tilting of the domains by the in-plane field leads to an effective reduction of the driving force on the DW. In the simplest model for this reduction one replaces $H_z \rightarrow H_z\sqrt{1-(H_x/H_{K,eff})^2}$ in the creep formula. Using this model with $H_{K,eff}$ as a fit parameter, we obtained fitted values for $H_{K,eff}$ that have the same order of magnitude as the ones that we measured directly, but were nonetheless unable to accurately describe all our measurements. This suggests that a more accurate calculation of the DW energy, including the tilting of domains by the in-plane field, is needed, which is the subject of future theoretical work.

In summary, we studied the DW velocity profiles of expanding bubbles in differently prepared Pt / Co / Pt samples under the application of in-plane fields. Specifically, we varied the Ar pressure during the deposition of the top Pt layer leading to a factor three increase in PMA between the lowest and highest pressure used. This indicates a large change in the structural quality between the samples. The velocity profiles of the expanding bubbels were successfully extracted. However, the results could not be described by a model that was used successfully before. We believe that the found results shed light on the understanding of the effective interfacial DMI in ultrathin films and will facilitate the quest to boost the DMI to stabilize chiral N\'{e}el DWs and create skyrmionic spin textures in thin film systems at room temperature.

\subsection{Acknowledgements}\label{SEC:ACK}
R.L. acknowledges support from the Netherlands Organization for Scientific Research (NWO-680-47-428). Solliance is acknowledged for funding the TEM facility. YY acknowledges support from the Foundation for Fundamental Research on Matter (FOM) which is part of NWO. AB acknowledges support by NanoNextNL, a Micro and Nanotechnology Consortium of the Government of the Netherlands and 130 partners.

\end{document}